# A two-scale model for sheared fault gouge: Competition between macroscopic disorder and local viscoplasticity

A.E. Elbanna

Department of Civil and Environmental Engineering,
2219 Newmark Civil Engineering Laboratory,
University of Illinois at Urbana-Champaign
Urbana, IL 61801, USA

J.M. Carlson

Department of Physics,
6123 Broida Hall,
University of California at Santa Barbara,
Santa Barbara, CA93106-9530, USA

**Abstract:** We develop a model for sheared gouge layers that accounts for the local increase in temperature at the grain contacts during sliding. We use the shear transformation zone (STZ) theory, a statistical thermodynamic theory, to describe irreversible macroscopic plastic deformations due to local rearrangements of the gouge particles. We track the temperature evolution at the grain contacts using a one dimensional heat diffusion equation. At low temperatures, the strength of the asperities is limited by the flow strength, as predicted by dislocation creep models. At high temperatures, some of the constituents of the grains may melt leading to the degradation of the asperity strength. Our model predicts a logarithmic rate dependence of the steady state shear stress in the quasi-static regime. In the dense flow regime the frictional strength decreases rapidly with increasing slip rate due to the effet of thermal softening at the granular interfaces. The transient response following a step in strain rate includes a direct effect and a following evolution effect, both of which depend on the magnitude and direction of the velocity step. In addition to frictional heat, the energy budget includes an additional energy sink representing the fraction of external work consumed in increasing local disorder. The model links low-speed and high-speed frictional response of gouge layers, and provides an essential ingredient for multiscale modeling of earthquake ruptures with enhanced coseismic weakening.

## I. Introduction

Understanding the dynamics of shear weakening in gouge layers, under a wide range of slip rates and confining normal stresses, is a fundamental and long-standing challenge in earthquake source physics. In the case of mature faults, that have accumulated hundreds of meters of slip throughout their active history, there is little doubt that shear deformation localizes to very thin zones typically less than 1mm wide [Chester, 1993; Chester and Chester, 1998; Chester et al., 2004; Noda and Shimamoto, 2005; Lockner et al., 2000; Ben-Zion and Sammis, 2003]. The zone which accommodates most of the slip within this layer can be even smaller. The absence of significant evidence of melting [Sibson, 1973; Lachenbruch, 1980], despite the extreme localization of strain, suggests that fault friction must be low during dynamic sliding [Kanamori and Heaton, 2000]. The fact that the static strength of fault gouge satisfies Byerlee's law (static friction coefficient 0.6-0.9), however, indicates that faults are statically strong. Understanding the mechanisms of dynamic weakening, from high static friction to low sliding friction, is essential for developing more accurate models of earthquake ruptures and scenarios for ground motion prediction [Rice, 1980, 2006; Heaton, 1990; Lapusta, 2000; Aagaard and Heaton, 2008; Ampuero and Ben-Zion, 2008; Noda et al., 2009, 2010].

In this paper, we present a two-scale model for dynamic weakening in gouge layers. Macroscopically, plastic strain results from the accumulation of local granular rearrangements. Microscopically, thermally activated viscoplastic processes at the grain contacts control sliding and force chain instabilities. The interplay between the two processes lead to non-monotonic strain rate dependence for the shear strength.

Irreversible local rearrangements of gouge particles is modeled using the shear transformation zone (STZ) theory, a continuum model of plastic deformation in amorphous solids that quantifies local configurational disorder [Falk and Langer, 1998]. The basic assumption in the theory is that plastic deformation occurs at rare non-interacting localized spots known as shear transformation zones (STZs). An internal state variable, the effective temperature, describes fluctuations in the configurational states of the granular material (i.e. a measure of local entropy), and controls the density of STZs [Langer, 2004; Haxton and Liu, 2007; Langer and Manning, 2008; Bouchbinder and Langer, 2009]. Effective temperature can be related to the system porosity [Lieou and Langer, 2012]. This approach coarse-grains granular simulations while retaining important physical concepts.

Gouge particles with dimensions of a micrometer and above are too big for thermal fluctuations to initiate transitions. Nonetheless, slip processes at the grain interfaces, such as dislocation glide and stable crack growth, are thermally activated [Chester, 1994, Rice et al., 2001; Noda, 2008]. In particular, the high confining pressure at depths relevant to earthquake nucleation and propagation lead to the prevalence of plastic conditions in the contact region. The local plastic rheology depends strongly on the local temperature. Moreover, at high slip rates, flash heating may occur leading to a significant degradation in the contact strength. Flash heating is the rapid increase in local temperature at the contact asperity due to heat generation due to frictional sliding at a rate higher than the heat diffusion rate [Rice, 2006]. By including the effect of local temperature changes on the evolution of the flow strength at the particle interfaces, it is possible to interpret the non-monotonic rate-dependent response of gouge layers observed experimentally at different slip rates and normal stresses [Chester, 1994; Blanbied et al., 1995; Sone and Shimamoto, 2009].

The primary result of this paper is the inclusion of a modified theory of flash heating within the framework of the STZ theory, providing a mechanism for (i) rate strengthening in the quasistatic regime of granular flow, and (ii) rate weakening in the dense regime of granular flow. This has important geophysical implications for nucleation, dynamic rupture, and energy partitioning during slip. The proposed theory is primarily relevant for small and moderate slips. For larger slips the rise in macroscopic temperature, under seismological conditions, will be large enough to cause macroscopic melting unless other weakening mechanisms such as pore fluid pressurization operate.

The remainder of the paper is organized as follows. In Section II we review the basic elements of the STZ theory. In Section III we discuss the model for local viscoplasticity at the grain contacts. In Sections IV and V we describe the procedure for calculating the contact temperature for both single and multiple contacts cases. In Section VI we discuss the parameter selection. In Section VII we investigate the predictions of the STZ theory for the steady state sliding shear stress as well as the transient behavior. We consider a wide range of strain rates. We also quantify the partition of dissipated energy between configurational and thermal components. We conclude in Section VIII by discussing implications of this model for dynamic rupture and gouge friction modeling.

## II. A review of the STZ theory

STZ theory is a non-equilibrium statistical thermodynamic framework for describing plastic deformations in amorphous materials by quantifying local disorder. It has been successfully applied to a variety of systems including granular fault gouge [Daub and Carlson, 2008; Daub et al., 2008, Daub and Carlson, 2010, Hermundstad et al., 2010], glassy materials [Falk and Langer, 1998, 2000; Manning et al., 2007, 2009], thin film lubricants [Lemaitre and Carlson, 2004], and hard spheres [Lieou and Langer, 2012]. In this section we review the basic assumptions and equations of this theory.

Particles in amorphous materials can move and rearrange in response to applied stress. The total shear deformation in any region can be approximated by two independent components; affine motion, in which particle displacements are homogeneous, and non-affine motion, in which particle displacements are inhomogeneous. Molecular dynamics simulations reveal that nonaffine motion is concentrated in localized regions, called shear transformation zones (STZs). These regions undergo configurational rearrangement by flipping between two bistable orientations, anti-aligned and aligned, under applied shear stress [Falk and Langer, 1998].

A single STZ event generates, on average, a fixed amount of local plastic strain $(\varepsilon)$ within the material. The macroscopic plastic strain is the cumulative result of many local events. Once flipped, STZs cannot further deform in the same direction. Instead, they are continuously created and destroyed in order to further accommodate plastic strain within the material.

The amount of configurational disorder in the system is characterized by a single state variable: the effective temperature $\chi$. The effective temperature is formally defined as the change in the system potential energy (or volume) per unit change in the system entropy [Bouchbinder and Langer, 2009; Lieou and Langer, 2012]. A fundamental result in the STZ theory is that the continuous creation and annihilation of the STZs drive their density $\Lambda$ toward a Boltzmann distribution $\exp(-1/\chi)$ [Langer, 2008; Langer and Manning, 2008; Bouchbinder and Langer, 2009].

The plastic strain rate $\dot{\gamma}$ is then given by:

$$\dot{\gamma} = (\varepsilon_o/\tau_o)\, R(s,\chi)\exp(-1/\chi), \qquad (1)$$

where $\varepsilon_o$ is the average plastic strain increment per STZ, $\tau_o$ is a characteristic time scale and $s$ is the shear stress. The rate at which STZs induce an infinitesimal plastic slip is given by the rate-switching function $R(s,\chi)$. The form of $R(s,\chi)$ is constrained by the second law of thermodynamics [Bouchbinder and Langer, 2009] and is given by:

$$R(s,\chi) = \begin{cases} (1 - s_o/s)\exp(s/s_c\chi), & \text{if } s > s_o; \\ 0, & \text{if } s \le s_o. \end{cases} \qquad (2)$$

The parameters $s_o$ and $s_c$ are two stress scales for the STZ system. The stress parameter $s_o$ is the minimum flow stress of the granular system. Thermal fluctuations are not sufficient to drive STZ reversal at the granular scale. Hence, $R(s,\chi)$ is nonzero only if $s > s_o$. Previously, it was shown that $s_c = p$ where $p$ is the pressure [Lieou and Langer, 2012].

The effective temperature evolves according to the following equation:

$$\dot{\chi} = \frac{s\dot{\gamma}}{s_o c_o}\left(1 - \frac{\chi}{\hat{\chi}(\dot{\gamma})}\right) + \frac{\partial}{\partial z} D\dot{\gamma}\frac{\partial \chi}{\partial z}. \tag{3}$$

Equation (3) states that only a fraction of the externally applied work rate $s\dot{\gamma}$ is dissipated to increasing $\chi$ as it is driven toward its steady-state value $\hat{\chi}(\dot{\gamma})$ [Langer and Manning, 2008]. This fraction is given by $1-\chi/\hat{\chi}$. The effective specific heat $c_o$ determines the amount of energy required to increase the effective temperature. The second term on the RHS of Eqn. (3) is effective only if $\chi$ is spatially heterogeneous. There, $D$ is the effective temperature diffusion coefficient and it scales with the square of particle size. Stability analysis [Manning et al., 2007] shows that the feedback between the strain rate and effective temperature may amplify spatial heterogeneities in the effective temperature and ultimately lead to shear banding. Homogeneous deformation corresponds to $\chi(z) =$ constant, where $z$ is the spatial coordinate across the gouge layer thickness.

The steady state effective temperature $\hat{\chi}(\dot{\gamma})$ satisfies the following condition: $d\hat{\chi}/dq \geq 0$ where $q = \dot{\gamma}\tau_o$ is the inertia number. Steady state simulations of glassy materials [Haxton et al., 2007; Langer and Manning, 2008] suggest that at high strain rates $\chi(\dot{\gamma})$ is given by:

$$\hat{\chi}(\dot{\gamma}) = \frac{A}{\ln(q_o / q)}, \tag{4}$$

where $q_o$ is the inertia number at which the effective temperature diverges. In this limit the system fluidizes and a solid-like description like the STZ theory is no longer applicable. The value of $q_o$ is estimated to be of the order 0.1-1 [Roux J.-N. and Chevoir F, 2005]. At low strain rates, the steady state effective temperature can be approximated by a constant value $\chi_o$ that is independent of strain rate. The rate parameter *A* determines whether the granular system is rate strengthening or rate weakening. Numerous experimental and numerical models for systems of hard and soft spheres under isothermal conditions predicts rate strengthening response [da Cruz et al., 2005 and references therein]. We choose a value of *A* consistent with these observations.

A typical strain rate dependence for the steady state shear stress of a system of hard spheres is shown in Fig. 1. No local plasticity at the grain level is considered and the system is assumed to deform under isothermal conditions. Three major flow regimes are identified in this case. In the quasi-static regime, prevailing at very low strain rates, the shear stress is almost independent of the strain rate. In the dense regime the shear stress increases with increasing strain rate. In the collisional flow regime, prevailing at high strain rates, kinetic energy due to collisional interaction between the particles becomes no longer negligible and STZ theory breaks down. In this limit, a hydrodynamic description is more relevant [Campbell, 1990].

The transition from quasistatic to dense flow depends on the properties of the granular system. This is shown in Fig. 2 where the value of the parameter $s_o$ is varied. The transition to dense flow occurs at a higher inertia number for systems with higher $s_o$. Figure 2 suggests that varying $s_o$ alone not only alters the strength at low strain rates but it also leads to different strengthening rates at intermediate and high strain rates.

The parameter $s_o$ plays a central role in STZ theory. If a stress lower than $s_o$ is applied to the system, the system undergoes a transient deformation but it eventually stops. In this limit, the force chains rearrange themselves and always find a stable configuration to resist the applied stress. For stresses higher than $s_o$, the force chains continuously collapse and reform but no stable configuration is achieved. Accordingly, a non-vanishing plastic strain rate is generated [Bouchbinder and Langer, 2009; Lieou and Langer, 2012]. The value of $s_o$ is a function of many system variables such as grain shape and surface roughness, physical chemistry of the particles, temperature and existence of fluids.

Different mechanisms contribute to flow resistance in granular systems. These include frictional sliding, particle interlocking, and rolling friction. If the grains are not perfectly spherical, as it is the case with most natural gouge particles, frictional sliding between particles dominate. In that limit it is possible to relate the value of $s_o$ to the local frictional strength, and hence viscoplastic processes, at the grain contacts. To induce local plastic slip and rearrangements, a force chain must buckle. This is possible if two particles in the chain slide relative to one another. This in turn requires that the local stress at the particle interfaces exceeds the frictional strength. The local stress and the macroscopic stress are related by:

$$sA = s_l A_r, \tag{5}$$

where $A$ is the apparent contact area (of order the particle cross sectional area), $A_r$ is the real contact area and $s_l$ is the local stress at the contact level. Similarly, one may assume that the parameter $s_o$ is related to the flow strength at the contact level by an analogous relation:

$$s_o A = s_{th} A_r, \tag{6}$$

where $s_{th}$ is the flow strength at the contact level. Equation (6) establishes a correspondence between the macroscopic yield stress $s_o$ and the local flow strength:

$$s_o = s_{th}\frac{A_r}{A}. \qquad (7)$$

This is an important constraint for grains with viscoplastic contacts. Combining Eqns. (5) and (6) we conclude that $s/s_o = s_l/s_{th}$. Accordingly, no persistent plastic deformation ($s > s_o$) is possible unless the local frictional strength is exceeded. We discuss this further, in the context of other contact models, in Sec V.

The ratio $A_r/A$ depends on pressure $s_c$ and the compressive strength of the grains. At the scale of microcontacts, the compressive strength is of the order of the indentation hardness $H$ of the grains [Rice, 2006; Noda, 2008]. It then follows from equilibrium of normal stresses that $A_r/A = s_c/H$. The indentation hardness depends on temperature; it decreases as the temperature increases. We ignore the modest evolution of hardness as a function of temperature in this study.

## III. Viscoplasticity at the grain contacts

To complete the temperature-dependent STZ model, we need to define an evolution law for the local flow strength at the grain contacts. Different possible mechanisms for plastic deformation at that scale exist. The dominant mechanism depends on many parameters such as temperature, pressure and grain size. On the macroscopic scale, the granular system is amorphous with no long range order associated with grain positions. Locally, however, the grains are crystalline solids susceptible to deformation by dislocation motion. For olivine, as an example, the dominant plasticity mechanism at high stresses and low temperatures is found to be dislocation glide [King and Marone, 2012]. As the temperature increases other mechanism may come into play such as diffusion creep in diffusion of vacancies lead to dislocations climbing over pinning sites in the lattice [King and Marone, 2012]. Eventually, the local temperature will be high enough to melt some of the minerals in the grains leading to a significant drop in the contact frictional strength [Rice, 2006].

To model dislocation glide, we follow Goetze [1978] and adopt the following formulation for the flow strength at the grain contacts:

$$s_{th} = \sigma_p \left[ 1 - \left( \frac{-RT}{\Delta H} \ln \frac{\dot{\gamma}}{B} \right)^{1/q} \right], \qquad (8)$$

where the Pierel's stress $\sigma_p$ = 8.5 GPa, the gas constant R= 8.314 J/(mol K), the activation enthalpy H = 5.4x$10^5$J/mol, the empirical constant B = 5.7 x $10^{11}$$s^{-1}$, and the exponent *q* is taken equal to 2.

At high slip rates, heat is generated at the contact surface faster than it can diffuse. It follows that the local temperature increases and may be high enough to generate a thin layer of molten material at the contact surface or at least melt some of the minerals in the grain. This layer lubricates the interface and reduces the local frictional strength to nearly zero. The occurrence of this flash melting depends on the grain size, the porosity of granular layer, the background temperature and the melting temperature of the grain constituents. There is also some evidence

that the strong degradation in the asperity strength may start at temperatures slightly lower than the overall melting point of the grain contact due to the melting of some of its minerals [Rice, 1999; Tullis and Goldsby, 2003; Rice, 2006; Beeler et al., 2008]. For this purpose, we assume the following empirical formula that connects the low temperature plastic strength of the asperity with the high temperature flow strength:

$$s_{th} = \sigma_p \left[ 1 - \left( \frac{-RT}{\Delta H} \ln \frac{\dot{\gamma}}{B} \right)^{1/q} \right] f\left(T, T_w\right), \qquad (9)$$

where $T_W$ is the characteristic weakening temperature of the grain mineral, and $T$ is the local contact temperature. It is given by $T = T_b + \Delta T$, where $T_b$ is the background temperature and $\Delta T$ is the local rise in temperature.

The function $f(T,T_w)$ is proposed to capture the degradation of the asperity contact strength at elevated temperature. We assume that $f(T,T_w) = 1$ for $T < T_w$. In this limit, Eqn. (9) coincides with Eqn. (8). On the other hand, for $T \gg T_w$, the strength function $f(T,T_w)$ should asymptotically approach zero. The rate of strength degradation depends on many factors such as the chemical environment, the grain metallurgy and the melting characteristics of the grain constituents. The specific form of $f(T,T_w)$ may be constrained from experiments or numerical simulations as we will discuss in Section VI. We emphasize, however, that $T$ in Eqn. (9) is the total temperature at the grain contact which is the sum of the background temperature and the rise in temperature due to the flash heating processes. It follows from Eqns. (7) and (9) that the minimum flow stress $s_o$ is given by:

$$s_o = \sigma_p \frac{s_c}{H}\left[1-\left(\frac{-RT}{\Delta H}\ln\frac{\dot{\gamma}}{B}\right)^{1/q}\right] f\left(T,T_w\right). \qquad (10)$$

Several models were proposed to describe viscoplasticity in crystalline materials. These include the simple Arrhenius-based activation model [Chester, 1994; Nakatani, 2001; Rice et al., 2001; Noda, 2008]; the Steinberg-Cochran-Guinan-Lund flow stress model [Steinberg et al., 1980] , the mechanical threshold stress flow model [Follansbe and Kocks, 1988]and Preston-Tonks-Wallace flow stress model [Preston et al., 2003]. In these last three models, the flow stress is expressed as a sum of two terms: an athemral component and thermal one. The thermal component of the flow stress is the product of two terms. The first one is the component of the flow stress due to thermally activated processes (e.g. dislocation motion). The second term takes the form $\mu(p,T)/\mu_o$ where $\mu(p,T)$ is the pressure and temperature dependent shear modulus and $\mu_o$ is a reference shear modulus. Equations (9) and (10) have a similar construction. There, the thermally activated component of the flow stress is given by the dislocation creep model (Eqn. (8)). Accordingly, we hypothesize that the function $f(T,T_w)$ may be reflecting thermal variation in the shear modulus of the region in the vicinity of the grain contacts. Constructing the function $f(T,T_w)$ from first principles is beyond the scope of this paper but we give an example for a plausible form in Section VI.

## IV. Single Contact Temperature Model

We assume that all contacts are initially at the background temperature $T_b$. The increase in temperature due to local shear heating at the grain contact is computed as follows. The heat generation rate per unit area during sliding is given by $s_l V = sA/A_r V^* = s\,H/s_c V^*$, where $V^*$ is the local slip rate. This heat source is assumed to be confined to the plane of contact. By convolving

the fundamental solution of the heat diffusion equation with this planar heat source [Appendix A], the temperature rise at the interface is given by:

$$\Delta T = \frac{1}{2\rho c\sqrt{\pi \alpha_{th}}} \int_{t}^{t+\Delta t} \frac{s(t')V^*(t')}{\sqrt{t-t'}} \frac{H}{s_c} dt' \quad . \tag{11}$$

The absolute temperature of the contact is then given by

$$T = T_b + \Delta T \ . \tag{12}$$

When the heat generation rate is constant, the maximum temperature rise occurs at the end of the contact lifetime, i.e. at $\Delta t = a/V^*$ where $a$ is the grain size. This maximum rise is given by:

$$\Delta T_{\max} = \frac{sV^*H}{2s_c\rho c\sqrt{\pi \alpha_{th}}} \sqrt{a/V^*} = \frac{sH}{2s_c\rho c} \frac{\sqrt{aV^*}}{\sqrt{\pi \alpha_{th}}} \ . \tag{13}$$

For general time-dependent heat sources, Eqn. (11) is integrated numerically. If the slip rate at the grain contact is not constant, the contact time is defined implicitly through the integral

$$\int_{t}^{t+\Delta t_c} V^* dt = a \ . \tag{14}$$

**V. Assembly of Contacts**

As the gouge layer is sheared, many grains are actively sliding past one another. These grain contacts are at different stages in their slip history; while two grains may be at the end of their contact lifetime, another pair may be just starting the process (Fig. 3). In the one dimensional idealization, adopted here, all STZs are arranged in series with respect to the direction of shear stress. That is, the macroscopic plastic strain rate is the sum of the individual microscopic plastic strain rates of the STZs. The shear stress for all STZs is the same, however. This is analogous to

a chain of grains that is sheared at one end. This has two implications. First, the slip rate at the grain contact is not, in general, equal to the imposed slip rate. Compatibility of macroscopic and microscopic deformations dictates that $N\Lambda V^{*} = \dot{\gamma} h = V$ , or equivalently:

$$V^{*} = \dot{\gamma} a/\Lambda \ . \tag{15}$$

Here, we made the assumptions that all grain contacts slide at the same rate $V_{stz}$ . Furthermore, we have let $a = h/N$ . Second, the minimum flow stress in the system $s_o$ is controlled by the sliding site with the highest temperature, or equivalently, the lowest strength. This latter conclusion follows from the chain analogy; under uniform stress a chain breaks at its weakest link. Equations (15) and (13) imply that the rise in temperature for gouge particles depends not only on the grain size but also on the degree of disorder in the system, as given by $\Lambda^{eq}$ . This is an important difference between sliding on a rock surface and sliding by shearing a gouge layer.

## VI. Parameter Selection

Two sets of parameters are used in the current model. The first set relates to the parameters of the STZ theory. The other set is related to the thermal properties of the gouge particles.

For the STZ parameters, we assume that the STZ strain $\varepsilon_o$ is of the order of unity [Lieou and Langer, 2012]. This corresponds to the assumption that during local rearrangements, a particle will move a distance that is approximately equal to its size. The inertia time scale $\tau = a/\left(s_c/\rho\right)^{0.5}$, where $\rho$ is the grain density, is the time required for a particle to move a distance comparable to its size $a$ under the influence of pressure $s_c$ [da Cruz et al., 2005; Lieou and Langer, 2012; Lieou et al., 2014]. This time scale $\tau$ has been found to appropriately describe grain dynamics

for a wide range of slip rates. The inertia number $q = \dot{\gamma}\tau$ is the ratio of the plastic strain rate $\dot{\gamma}$ to the strain rate associated with the inertial motion $1/\dot{\tau}$. In applications relevant to seismology, the pressure may be as large as a few hundred MPa. We take $s_c$ in the subsequent calculations to be equal to 75 MPa unless otherwise stated. The critical inertia number $q_o$, at which the disorder temperature diverges and the inertial flow commences, is assumed to be 0.1-1 [da Cruz et al., 2005; Roux J.-N. and Chevoir F, 2005; Job et al., 2006]. Previously [Langer and Manning, 2007.] it was shown that in the limit of small strain rates, the steady state value of the disorder temperature approaches a constant value $\chi_o$ that is independent of strain rate. The specific value depends on the nature of the system of particles (e.g. particle shapes and grain size distribution). Numerical simulations of systems of hard spheres suggest that $\chi_o \approx 0.2$ [Haxton et al., 2011; Lieou and Langer, 2012.]. In subsequent calculations, we vary $\chi_o$ between 0.15 and 0.30. Higher values of $\chi_o$ correspond to less compacted states.

For the properties of the gouge particles, we assume the following nominal values: density $\rho = 2700$ kg/m$^3$, volumetric heat capacity $\rho c = 2.7\times10^6$ J/m$^3$k, coefficient of thermal diffusivity $\alpha_{th} = 10^{-6}$ m$^2$/s, and hardness $H = 6$ GPa [Rice, 2006; Noda, 2008]. These values weakly depend on temperature.

**VII. Results**

To characterize the influence of local heating on the frictional response of the gouge layer, we investigate both the steady state and transient behavior. We also examine the energy partitioning between frictional heat and configurational work.

### VII.1. Steady state response

If the gouge layer is sheared for a sufficiently long time under constant slip rate, steady state conditions will prevail. That is, no further evolution in the shear stress, plastic strain rate or disorder occurs. Mathematically, that is equivalent to setting $\dot{\gamma} = V/h$, $\chi = \hat{\chi}$ and $s = s_{ss}$ in Eqn. (1). The minimum flow stress $s_o$ is calculated based on Eqn. (10) with $T_{ss} = T_b + \Delta T_{ss}$ (Eqn. (12))where $\Delta T_{ss}$ is the rise in contact temperature at steady state due to local heating. This latter quantity can be estimated from Eqn. (13) by setting $V^* = V_{stz} = \frac{\dot{\gamma} a}{\Lambda} = \frac{V}{h} a \exp(1/\hat{\chi})$. It follows that:

$$\Delta T_{ss} = \frac{sHa}{2 s_c \rho c} \sqrt{\frac{\exp(1/\hat{\chi})}{\pi \alpha_{th} h}} \sqrt{V} \quad . \qquad (16)$$

Here we have made use of the relation $\Lambda = \exp(-1/\chi)$. The steady state shear stress is then determined by solving the following equation simultaneously with Eqn. (16)

$$\frac{V}{h} = (\varepsilon_o/\tau_o)\left(1 - \frac{s_o(T_{ss})}{s_{ss}}\right)\exp\left(\frac{s_{ss}}{s_c \hat{\chi}}\right)\exp(-1/\hat{\chi}). \qquad (17)$$

The minimum flow stress $s_o$ is calculated using Eqn. (10), and the form of $f(T, T_w)$ is constrained by experimental measurements.

As an example, we consider the series of high speed frictional experiments of Sone and Shimaoto [2009]. In these experiments, a layer of fault gouge 1mm thick is sheared at different slip rates in the range of 0.1-2.1 m/s under 0.56 MPa normal stress. They reported that the steady state shear stress varies exponentially with the imposed slip rate. This is represented by the discrete points in Fig. 4.

To fit the experimental observations, we solve Eqns. (10), (16) and (17). For this purpose, we make the simplified assumption that the steady state effective temperature is constant and independent of strain rate, i.e. $\hat{\chi} = \chi_o$. This may be justified as follows. The fluidizing strain rate $\dot{\gamma}_c$ corresponding to the experimental conditions discussed here is given by $q_o/\tau$ where $q_o = 1$ and $\tau = a/\sqrt{s_c/\rho} = 6.94\times 0^{-7}\, s$ (assuming average grain size $a = 10\mu m$). This yields $\dot{\gamma}_c = 1.44\times 10^{6}\ s^{-1}$ which is almost three orders of magnitude larger than the highest macroscopic strain rate reported in the experiment $(2100\ s^{-1})$. For strain rates that are much lower than the fluidizing strain rates, previous work [Langer and Manning, 2007] suggests that the steady state effective temperature is well approximated by its low strain rate limit. We adopt this simplification here and assume $\chi_o = 0.15$. The weakening temperature $T_w$ varies depending on the material composition of the grains. It ranges from $750\,^{o}C$ for biotite [Rice, 2006] to approximately $1200\,^{o}C$ for silicates. Here, we choose $T_w = 1300\ K$.

For low strain rates, the increase in contact temperature is miniscule (a few degrees) [Mair and Marone, 2000; Mair et al., 2006]. In this limit, our theory predicts that the steady state response is rate strengthening (Fig. 3). In Section II we showed that STZ theory predicts rate independent behavior for sheared granular material at low strain rates, and this prediction is consistent with numerical simulation of soft and hard spheres [e.g. da Cruz et al., 2005]. The rate strengthening response shown here has been reported in frictional experiments on fault gouge [e.g. Chester 1994; Blanpied et al., 1995] and has been explained within the framework of rate and state friction [Dieterich, 1979; Ruina, 1980] which was developed primarily for sliding on rock surfaces. We attribute this behavior to the rate dependence of the contact strength (Eqn. (8)) that overtakes the weakening effect due to temperature changes. Figure (4) suggests that for slip rates

smaller than 1mm/s the steady state friction coefficient depends logarithmically on the slip rate and that $\partial\mu_{ss}/\partial\log V = 0.0075$. Similar values have been documented experimentally [Dieterich, 1979; Ruina, 1980; Chester, 1994; Blanpied et al;, 1995; Marone, 1997].

To match the behavior at high strain rates, corresponding to $V > 0.1$ m/s, the functional form of $f(T, T_w)$ must be determined. We found the following form to yield the best results:

$$f(T, T_w) = C\exp\left(-r\left(\frac{T - T_b}{T_w - T_b}\right)^2\right), \tag{18}$$

where $T_b = 300$ K, $r = 2$ and $C = \exp(r)$. Expanding the exponential function in Eqn. (18) and defining $T_* = (T - T_b)/(T_w - T_b)$ we obtain the following approximation:

$$f \approx 1 - rT_*^2, \tag{19}$$

In this limit, we recover the temperature-dependent term in the Johnson-Cook flow stress model with $m = 2$ [Johnson and Cook, 1983]. This approximation is valid for $T$ in the vicinity of $T_w$.

The predictions of the model fit the experimental measurements very well for slip rates ranging between 0.1 m/s and 1.25 m/s. For higher slip rates, the friction coefficients reported experimentally are lower than what the model predicts. We hypothesize that this discrepancy may be attributed to strain localization which is enhanced at higher slip rates. In the current model, the strain rates are computed assuming a shear zone thickness of 1mm. Sone and Shimamoto [2009] observed that the strain may localize to bands 100-150 $\mu m$ thick. The strain rate in the shear band is usually higher than the average strain rate [Manning et al., 2007]. The shear band thickness broadens with time [Manning et al., 2007; Lieou et al., 2014], nonetheless,

and its dynamics depend on the imposed strain rate. We expect strain localization to be enhanced at higher slip rates. This in turn leads to larger increases in local temperature and lower flow strength. Thus, we conjecture that modeling strain localization will result in a better quantitative fit for the data at higher slip rates. This will be the subject of a future investigation.

### VII.2. Transient Response

To examine the full response of the sheared gouge layer, including direct and evolution effects, we integrate the STZ equations of motion, Eqns. (1) and (3), coupled with the equation for contact temperature (Eqn. (11)). To complete the dynamical description of the system, we augment the STZ equations by an equation for the evolution of the shear stress:

$$\dot{s} = G\left(\frac{V}{h} - \dot{\gamma}\right), \tag{20}$$

Where $G$ is the shear stiffness of the system per unit length and $V/h$ is the imposed strain rate. We assume $G = 30$ GPa and consider high strain rates in the range of $10^3$ to $10^4 s^{-1}$. We assume $s_c = 75$ MPa, $a = 1$ µm, $T_b = 273$ K, $T_w = 1000$ K, and $\chi_o = 0.27$. These values lead to reasonable computational cost and are within the range of physically plausible limits for fault gouge. All other parameters are taken as in the previous sections.

#### VII.2.1. Velocity stepping experiment

Figures 5a shows the result for a pair of velocity stepping numerical experiments between strain rates $10^3$ and $2\times10^3\ s^{-1}$. The inserts expand the scales to visualize the shear stress variations right after the upward and downward velocity steps. Several remarks follow from this figure. First, the response is asymmetric for the upward and downward steps. The downward direct effect is slightly larger with a magnitude of 0.154 MPa whereas the magnitude of the upward

direct effect is 0.144 MPa. Second, the evolution of friction following the downward step is steeper than the one following the upward step. Third, following a mathematical step in imposed slip rate, the frictional stress does not follow with an infinite slope as predicted for bare rock surfaces in contact [Noda, 2008]. Rather, it gradually evolves, especially in the downward velocity step case.

The previous observations continue to hold for larger changes in strain rates. Figure (5b) shows the results for a velocity stepping numerical experiment between $10^3$ and $10^4\ s^{-1}$. The upward direct effect is equal to 0.5 MPa while the downward direct effect is equal to 0.64 MPa, compared to 0.144 MPa and 0.154 MPa in the smaller step magnitude experiments, respectively. These results suggest that the normalized direct effect, $\partial s / \partial \left( s_c \log \left( V_2 / V_1 \right) \right)$, increases with the magnitude ratio of the velocity step $V_2 / V_1$.

The asymmetry between the upward and downward direct effect, which increases as the magnitude of the strain rate step increases, may be explained by the thermo-mechanical coupling in our model. As the system is sheared, grains come into and out of contact continuously. Following a velocity step, regardless of its direction, the contact temperature will initially increase. We show in Figs (5c) and (5d) an example for the temperature evolution corresponding to the stress-strain curve in Fig. (5a). In case of an upward velocity step, the contact temperature continues to rise (Fig. 5c). For a downward velocity step, however, the contact temperature will reach a maximum and then decreases due to the reduced heat generation rate (Fig. 5d). This difference in the evolution history of the contact temperature results in different evolution for $s_o$ and consequently different variation in the shear stress. A similar trend was observed in the temperature variations corresponding to the stress-strain curve shown in Fig. (5b).

A common feature in Figs. 5a and 5b is that the evolution of stress as a function of strain, following an upward step in velocity, is non-monotonic. The stress first decreases until it reaches a minimum and then it increases again towards a steady state. This is explained by the evolution of the local temperature at the grain contacts. Following an upward step in velocity, the contact temperature first increases (see Fig. 5c for example). Diffusion effects become important as slip accumulates. Accordingly, the temperature will reach a maximum and then gradually decreases due to heat diffusion. This temperature trend causes the flow stress to first decrease and then increase. The shear stress follows the flow stress trend (Fig. 5a).

**VII.2.2 Velocity ramps experiment**

The high speed frictional experiments of Sone and Shimamoto [2009] were unique in their ability to test the frictional response under decelerating and accelerating velocity ramps. Figure 6 shows the results of an analogous numerical experiment in which the imposed strain rate increases linearly between $10^3$ and $10^4 s^{-1}$ over a strain of 0.005 and then decreases again to $10^3 s^{-1}$ over a strain of 0.015, after which it remains constant. In this case, it is observed that the upward direct effect is reduced significantly compared to Fig. 6. During the upward velocity ramp, the shear stress gradually decreases. The minimum shear stress value is comparable to the minimum value in Fig. 5. However, it does not occur at the strain corresponding to the maximum strain rate. The shear stress continues to decrease during a part of the downward velocity ramp. This is attributed to the reduced value of the minimum flow stress that resulted from the increase in temperature during the upward velocity ramp. Eventually, the shear stress starts to increase

while the imposed strain rate is decreasing. The cooling of the asperities due to the reduced heat flux leads to an increase in $s_o$ which in turn results in shear strengthening.

In Fig. 6b, we show experimental results from Sone and Shimamoto (2009) for a velocity ramping experiment. In their experiment, a shear band of 100-150 µm was observed. Assuming that the strain rate is totally accommodated by deformations within the band, a ramp in the imposed slip rate between 0.1 and 1.0 m/s corresponds to a ramp in strain rates in the shear band from $10^3$ to $10^4$ $s^{-1}$ (assuming shear band thickness 100 um). This is the range of strain rates we tested numerically in Fig. 7a. The initial frictional strengthening observed experimentally is due to starting the experiment from zero velocity which requires overcoming the "static friction" in order for the plastic strain rate, and subsequent softening, to start to accumulate. In the numerical model, however, the sample is sheared initially at $10^3$ $s^{-1}$, and by the time the velocity ramp is applied the system has significantly deformed. Another important difference between the experiment and the numerical model is the pressure scale. The normal stress applied experimentally is of the order of 0.56 MPa which is 150 times smaller than what is used in the numerical simulations. This leads to experimentally longer evolution length and time scales. Nonetheless, the model predictions and the experimental observations show good qualitative agreement in the trend for stress evolution.

### VII.3. Energy partitioning

In this section we quantify how the energy is partitioned in the gouge layer between its different components. As the gouge layer is sheared, energy is dissipated as radiated energy, frictional heat, and increases in local disorder. The total dissipation is given by:

$$E_T = \int s(u)du \,. \tag{20}$$

The configurational energy $E_c$ is the fraction of energy required to change the effective temperature and increase local disorder. The configurational energy per unit area is given by:

$$E_c = h\int c_o s_c d\chi = \int \left[ hc_o s_c \, d\chi / du \right] du. \tag{21}$$

In this paper we do not consider inertial effects or energy lost to radiation. We thus attribute the remaining dissipation to thermal heating. The heat dissipated per unit area $E_f$ is the difference between the total and configurational energy per unit area,

$$E_f = E_T - E_c = \int \left[ s(u) - hc_o s_c \, d\chi / du \right] du. \tag{22}$$

An example of the partitioning between these two modes of energy dissipation is shown in Fig. 8. The amount of energy dissipated per unit area as frictional heat and configurational energy are represented by the yellow and red areas respectively. The upper curve in both figures represents the actual shear stress in the system. The boundary between the red and the yellow regions represents the frictional heat stress $s_f$. This is defined as fraction of shear stress contributing to heat generation. From Eqn. [22] this fraction is given by:

$$s_f = s(u) - hc_o s_c \, d\chi / du. \tag{23}$$

We can immediately see from Fig. 7 that configurational energy is not be a negligible part of the energy budget. In this particular case, nearly 10% of the dissipated energy is consumed in increasing local disorder. This percentage varies depending on pressure, the initial value of the effective temperature (which reflects how loose or compact the sample initially is), the value of steady state effective temperature, and the magnitude of slip. If the system is sheared long

enough to reach configurational steady state, there will be no further increase in the local disorder and the configurational energy will diminish. In that limit, all work as dissipates as heat.

## VIII. Discussion

Earthquakes are frictional instabilities which occur because fault strength weakens with increasing slip or slip rate [Rice, 2006]. Field observations suggest that slip in individual events on mature faults occurs primarily within a thin shear zone, <1–5 mm, that occurs inside a finely granulated fault zone [Chester and Chester, 1994]. In absence of a strong weakening mechanism, temperature rise would lead to widespread melting. Nonetheless most exhumed faults shows a lack of evidence for the existence of pseudotachylyte that would be left from rapid recooling of the molten rocks. Relevant weakening processes in crustal events are likely to be activated by thermal processes during dynamic slip. Several processes have been proposed such as (1) thermal pressurization of pore fluids, and (2) flash melting at highly stressed frictional microcontacts [Rice, 2006]. In this paper, we implement temperature-dependent viscoplasticity and the theory of flash heating within the framework of Shear transformation Zone (STZ) theory to predict the frictional response of sheared gouge layers at different slip rates and confining normal stresses.

Important differences exist between sliding on bare rock surfaces and shearing a layer of granular materials. Mainly, granular particles possess extra degrees of freedom compared to asperities on rock surfaces. Grains have both translational and rotational degrees of freedom and the plastic strain rate is partitioned between these different modes. This further enables grains to rearrange and to accumulate plastic deformation through this process. This flexibility is not available for rock asperities.

For rock surfaces sliding past one another, the asperities on the contact surfaces are arranged in parallel with the shear force. As a result, it is natural to assume that (1) the slip rate at all active asperities is equal to the imposed slip rate, and (2) the applied shear force is the sum of the local shear forces at the asperities. It follows that the strength of the interface is the summed resistance of all contacts and is not governed by the weakest one. In this case, the strength is determined by the average contact temperature and not the maximum temperature [Noda, 2008].

In sheared granular materials the opposite situation occurs. STZs are arranged in series with the shearing force, analogous to a chain of beads that is loaded at one end (Fig. 3). The shear stress may be assumed to be constant across the granular layer so that all STZs are subject to the same stress. The imposed slip rate, however, is partitioned among the active STZs. Accordingly, the local slip rate at an STZ site may be significantly different from the imposed slip rate depending on the STZ density. The larger the number of STZs the smaller the local slip rate compared to the imposed slip rate. Another consequence is that the strength of the system is governed by the strength of the "weakest" STZ. Using the chain analogy and under uniform stress, the strength of the chain is governed by its weakest link. In the context of the current paper, the weakest link corresponds to the sliding site with the smallest local friction or the highest contact temperature.

Other theories for temperature dependent weakening mechanisms in gouge have been proposed earlier. Chester (1994) proposed a state-variable constitutive relation, within Dieterich-Ruina rate and state friction framework, which describes the dependence of friction on temperature near steady state conditions. Their approach is suitable for very low slip rates (μm/s) and is based on the assumption that the micromechanisms of friction are thermally activated and follow an Arrhenius relationship. Here, we consider the general umbrella of viscoplastic processes that may occur at the grain contacts at different temperatures and slip rates. By incorporating

viscoplasticity with the STZ theory, which takes into account the amorphous nature of the granular layer, we connect the local processes at the grain contacts with the macroscopic plastic strain accumulated by particles slip and rearrangement. The relevant temperature for the activation of the viscoplastic processes is the absolute local temperature at the grain contact. This is the sum of the background macroscopic temperature and the local temperature rise due to flash heating processes. We were able to numerically calculate the changes in temperature at the grain level for different slip rates and to quantify the interaction between thermally activated viscoplasticity and granular dynamics. We showed that this interaction may lead to a non-monotonic granular rheology. Incorporating these thermally activated processes within the STZ framework will allow us to explore the effect of temperature changes on processes like shear banding which are not directly resolved by the other approaches.

In this paper, we assumed that the background temperature is kept constant by continuous cooling. In real faults, this condition is not satisfied in general. Sliding for long distances will inevitably cause an increase in the macroscopic temperature. This will require solving the heat diffusion equation for the background temperature with the heat source term corresponding to the macroscopic heat generation rate $s\dot{\gamma}$. In this case the absolute temperature at the grain contact is calculated by adding the local rise in temperature to the evolving background temperature. The higher the background temperature, the smaller the increase in local temperature required to initiate strong rate weakening. We thus expect that modeling the changes in macroscopic temperature will modify the slip rate at which significant weakening is observed.

The form of the flow rule used here to describe dislocation glide was developed specifically for Olivine [Goetze, 1978]. It may be regarded as a special case of the more generalized mechanical stress threshold flow model [Follansbe and Kocks, 1983]. At low slip rates, the adopted flow rule

predicts changes in frictional strength similar to those observed experimentally (See Fig. 3). At high slip rates, the influence of the thermally activated part of the flow rule diminishes as the strength is controlled more by the flash heating processes and the function $f(T,T_w)$. We expect qualitatively similar behavior to be obtained if a simplified Arrhenius relationship [Chester, 1994; Noda, 2008] is used for the flow rule. The advantage of the mechanical stress threshold model is that it is more flexible and enables the inclusion of microstructure damage evolution at the asperity level. This is relevant for further improvement in the theory.

Granular dynamics is usually described as rate insensitive at low strain rates and rate strengthening at high strain rates. This has been reported in numerous numerical and experimental observations [e.g. da Cruz et al., 2005]. Fault gouge, however, exhibit viscoplastic properties and was shown experimentally to exhibit both rate strengthening and rate weakening response [Chester, 1994; Blanpied et al., 1995; Sone and Shimamoto, 2009]. Our modified version of STZ theory with temperature-dependent viscoplastic interactions predicts a modest rate strengthening at low strain rates. At high strain rates, the contact temperature rises significantly. This leads to strong rate weakening that mimicks the exponential degradation in strength reported in Sone and Shimamoto [2009].

In shearing dry granular layers, dissipated energy is partitioned between frictional heat and configurational energy. The former is responsible for increasing the layer temperature while the latter facilitates local disorder [Hermundstadt et al., 2010]. Thermal weakening reduces the total amount of dissipated energy compared to cases when the strength is modeled as a temperature independent property by reducing the value of the sliding stress at any given strain. Moreover, we have shown that a portion of the total energy, estimated here to be of the order of 10%, is

expended in increasing local disorder. This further reduces the amount of energy dissipated by heat.

Depending on the strain rate and grain size distribution, strain may localize to a band that is less than 1mm thick [Sone and Shimamoto, 2009]. Shear banding is as an additional weakening mechanism that affects both the level of the sliding shear stress and the slip weakening distance. STZ theory provides a powerful tool for resolving strain localization at different strain rates [Daub et al., 2008; Manning et al., 2009; Hermundstad et al., 2010]. We expect thermal weakening to enhance the localization process and to affect the dynamics of growth of the shear band.

In continuum systems, a fraction of total energy is dissipated as radiated energy. This additional energy sink was not considered in this paper. We expect systems which weaken rapidly, as a result of thermal weakening, to radiate more energy for the same amount of slip. This in turn will affect the resulting ground motion and in particular its high frequency content..

Seismic inversions in a number of events, such as the Tohoku earthquake [Mori, 2011; Romano et al., 2012], suggest that dynamic rupture propagated for significant distances in what had been identified as rate strengthening regions. Here we considered a rate strengthening granular material as a basic model but showed that thermal weakening, due to local heating of grain contacts, may provide a mechanism by which a material, that is rate strengthening at low slip rates, significantly weakens at high slip rates. This implies that rate strengthening regions may not always impede rupture propagation. Accordingly, on a fault with both rate strengthening and rate weakening regions, limiting the expected maximum size of an earthquake to the length of

the rate weakening zone only could be a dangerous underestimation with serious implications for seismic hazard.

A controlling parameter for the strength of gouge particles with viscoplastic properties is the maximum increase in temperature at the grain contact. The theoretical formulation presented in this paper suggest that this maximum temperature does not depend only on the grain size but also on the state of disorder of the system (see, e.g., Eqn. 16). Compacted systems are more susceptible to higher increases in temperature. This dependence on the degree of disorder, or the way of sample preparation in terms of looseness or compactness, has no analogy in the case of sliding on bare rock surfaces and is an important characteristic of granular systems

As the material is sheared, two competing mechanisms control the temperature variations; the rate of heat generation and the rate of thermal diffusion. For high slip rates, the former surpasses the latter, and heat is localized, leading to temperature rise and reduction in contact strength. If the slip rate decreases, the effect of thermal diffusion will eventually dominate leading to a reduction in the temperature. This increases the contact strength and leads to re-strengthening. Hence, the temperature dependence of the contact strength allows for both rapid weakening and healing of the sheared gouge layer. This provides a framework for the analysis of constitutive response for different loading rate scenarios, including velocity stepping and ramping experiments, as discussed here, as well as slide-hold-slide tests. In particular, accounting for flash processes at grain contacts may be an important ingredient in understanding the difference in healing rates following low and high speed frictional experiments.

## Acknowledgements

The authors wish to thank James Langer, and Ralph Archuleta for the helpful discussions. A.E. Elbanna is grateful for H. Sone and H. Shimamoto for sharing their data and insights into the experimental implications. This research was funded by NSF/USGS Southern California Earthquake Center, funded by NSF 489 Cooperative Agreement EAR0529922 and USGS Cooperative Agreement 07HQAG0008, and by Office of Naval Research MURI grants N000140810747 (http://www07.grants.gov/search/search. do?oppId = 42747&mode = VIEW) and 0001408WR20242.

## Appendix: Estimation of the local increase in contact temperature

In this appendix we derive Eqn. (11) and explain the numerical integration approach in the case of multiple STZs.

The fundamental solution of the heat diffusion equation in one dimensional is given by:

$$\Phi\left(x,t\right)=\frac{1}{\sqrt{4\pi\alpha t}}\exp\left(-\frac{x^2}{4\alpha t}\right), \tag{A.1}$$

where $\alpha$ is the thermal diffusion coefficient. For an arbitrary heat source $g(x,t)$, the temperature distribution is obtained by convolving the heat source with the fundamental solution:

$$T(x,t)=\Phi\left(x,t\right)*\mathrm{g(x,t)}=\int_0^t\int_{-\infty}^{\infty}\frac{1}{\sqrt{4\pi\alpha\left(t-t'\right)}}\exp\left(-\frac{\left(x-y\right)^2}{4\alpha\left(t-t'\right)}\right)g\left(y,t'\right)dydt'\,. \tag{A.2}$$

For a planar heat source localized at the grains contact surface $g(x,t)=g_o\left(t\right)\delta\left(x\right)$. It follows from the properties of delta function that:

$$T(x,t)=\int_0^t\frac{1}{\sqrt{4\pi\alpha\left(t-t'\right)}}\exp\left(-\frac{x^2}{4\alpha\left(t-t'\right)}\right)g_o\left(t'\right)dt'\,. \tag{A.3}$$

The maximum rise in temperature occurs at the contact; that is at $x=0$. Hence,

$$T(0,t)=\int_0^t\frac{g_o\left(t'\right)}{\sqrt{4\pi\alpha\left(t-t'\right)}}dt' \tag{A.4}$$

If we take $g_o(t) = s(t)V(t)/\rho c$ , where $\rho c$ is the volumetric heat capacity, Eqn. (A.4) will reduce to Eqn. (11) in the text.

Equation (A.4) represents a convolution in time. At a given time *t*, the rise in temperature is due to the combined effects of all heat sources acting in the time period $0 \le t' < t$. Explicitly,

$$T(0,t) = \frac{g_o(0)}{\sqrt{4\pi\alpha(t-0)}}\Delta t_1 + \frac{g_o(\Delta t_1)}{\sqrt{4\pi\alpha(t-\Delta t_1)}}\Delta t_2 + \frac{g_o(\Delta t_1 + \Delta t_2)}{\sqrt{4\pi\alpha(t-\Delta t_1-\Delta t_2)}}\Delta t_3 + \ldots + \frac{g_o\left(\sum_{i=0}^{n-1}\Delta t_i\right)}{\sqrt{4\pi\alpha\left(t-\sum_{i=0}^{n-1}\Delta t_i\right)}}\Delta t_n. \tag{A.5}$$

In Eqn. (A.5), $\Delta t_i$ is the time duration for which the heat source $g_o(\Delta t_{i-1})$ is active, and we let $g(\Delta t_0) = g_o(0)$. We use the fact that $\sum_{i=0}^{n}\Delta t_i = t$. In this paper $\Delta t_i$ is less than or equal to the time step required for integrating the STZ equations.

In deriving Eqn. (A.4) the limits of integration were taken as 0 and *t*. A contact surface, however, does not sense the effect of heat sources that operated before the contact exists. Furthermore, it will not increase in temperature after the grains lose contact. Hence, the limits of integration in this case correspond to the beginning and end of the contact lifetime. In case the contact slip rate is not constant, Eqn. (14) must be used to determine the contact lifetime.

We calculate the increase in temperature for all STZs. We record the highest contact temperature only since it controls the flow stress in the system. Once the STZ with the highest temperature completes its slip, we consider the next STZ in the hierarchy, i.e. the STZ with the second highest temperature. We track the temperature evolution for this STZ until it completes its slip. We repeat this for all remaining STZs.

**Figures**

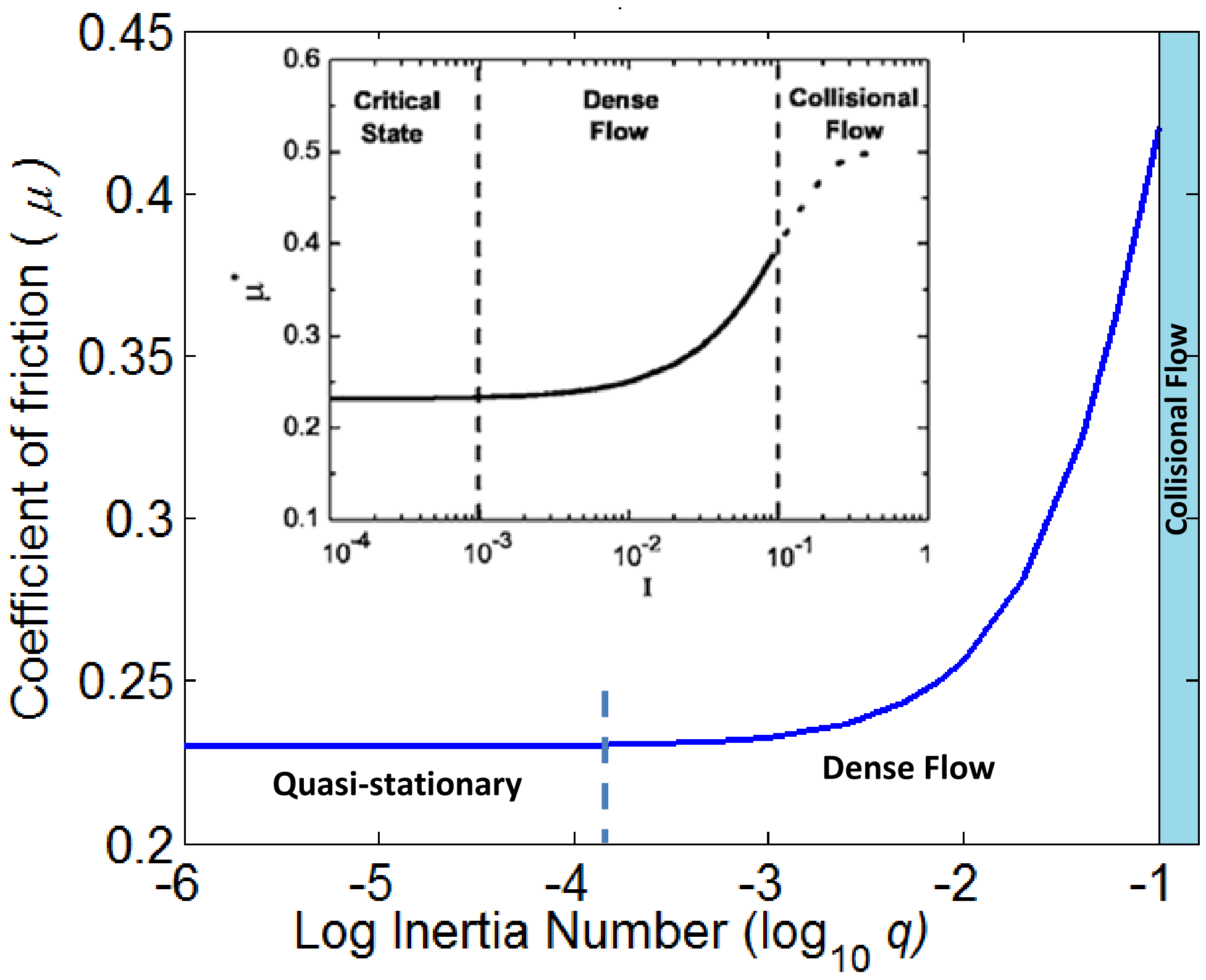

Fig. 1: Predictions of STZ theory for the rheology of sheared granular system with no local viscoplasticity at grain contacts. Insert shows results from molecular dynamic simulations [da Cruz et al., 2005, reprinted with permission]. [Simulations are shown for $s_o/s_c = 0.23$, $\varepsilon = 1$, $\chi_o = 0.2$].

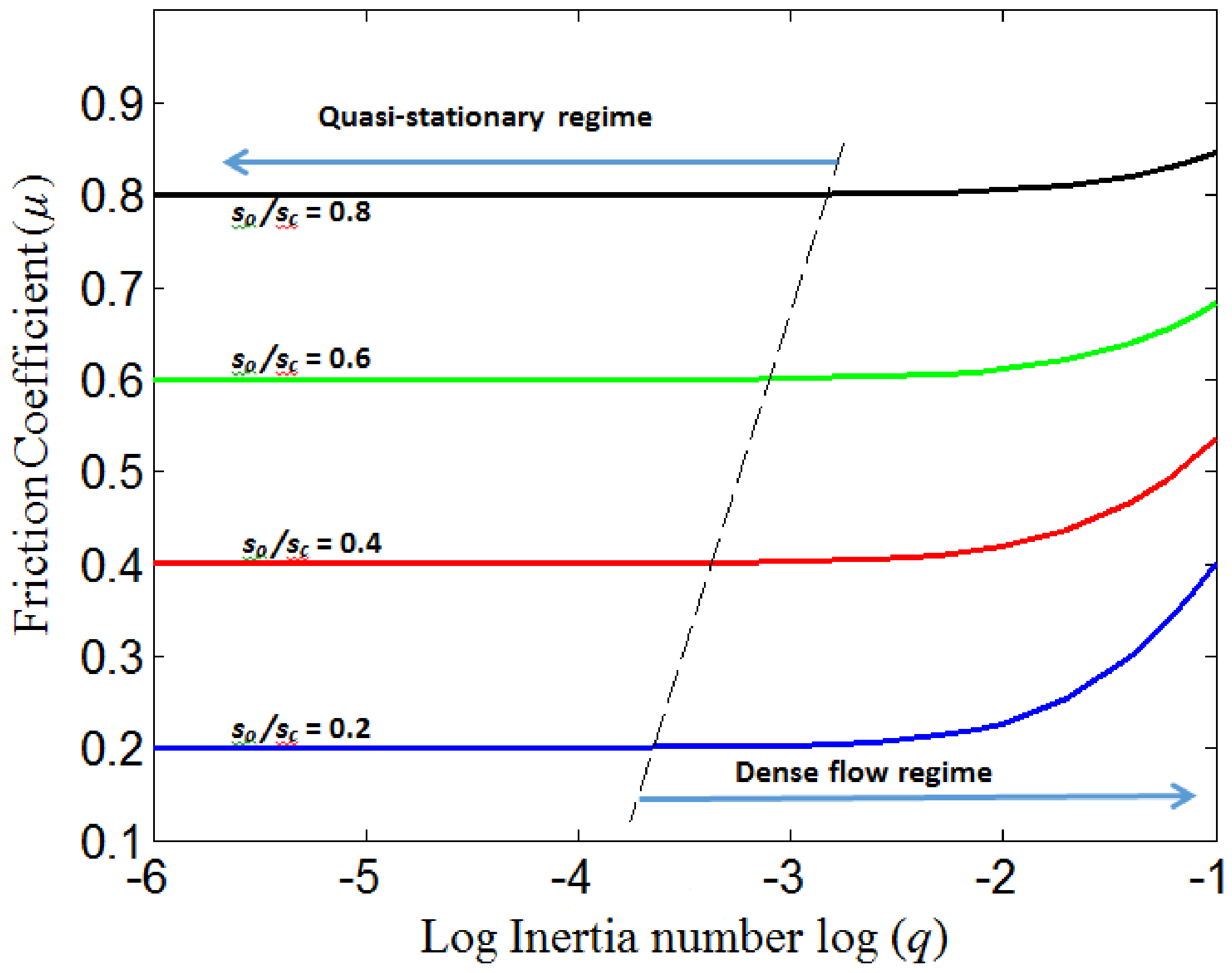

Fig. 2: Variation of coefficient of friction $\mu$ as a function of inertia number $q$ for different values of $s_o$ normalized by the pressure $s_c$. Systems with higher $s_o$ exhibit a transition from quasi-static to dense flow at higher strain rates. This transition is traced approximately by the dashed curve. The dense flow regime also has different strengthening rates for different values of $s_o$. [The same parameters are used here as in Fig. 1, but with different $s_o/s_c$ ratios.]

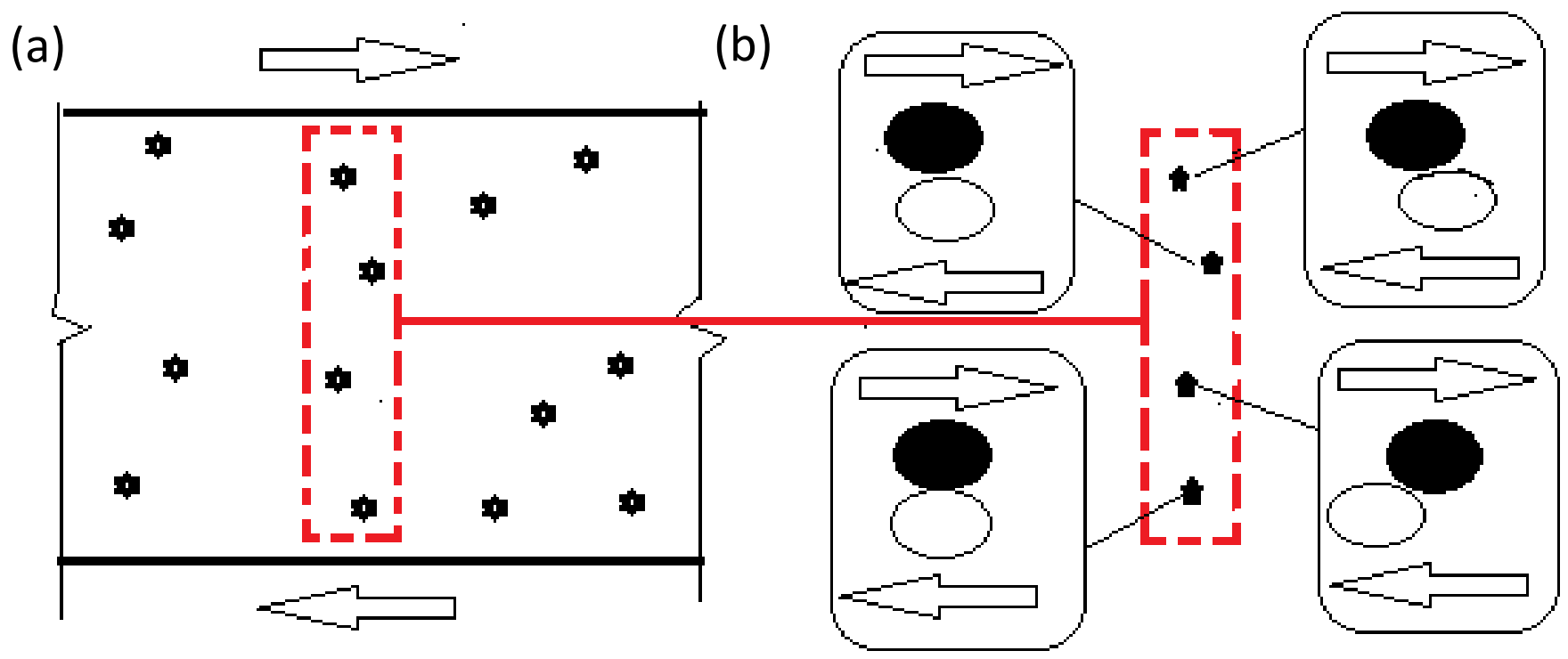

Fig. 3: A schematic for a 2D sheared granular layer with local plastic zones represented by stars. (a) The 2D representation. STZs are distributed along the depth as well as the width of the sample. The arrow represents the direction of shear. (b) The idealized 1D model assuming an infinite strip. Local plastic deformation is accommodated by inter particle slip as shown in the oval inserts. At any given instant, different particles are at different stages of their slip history.

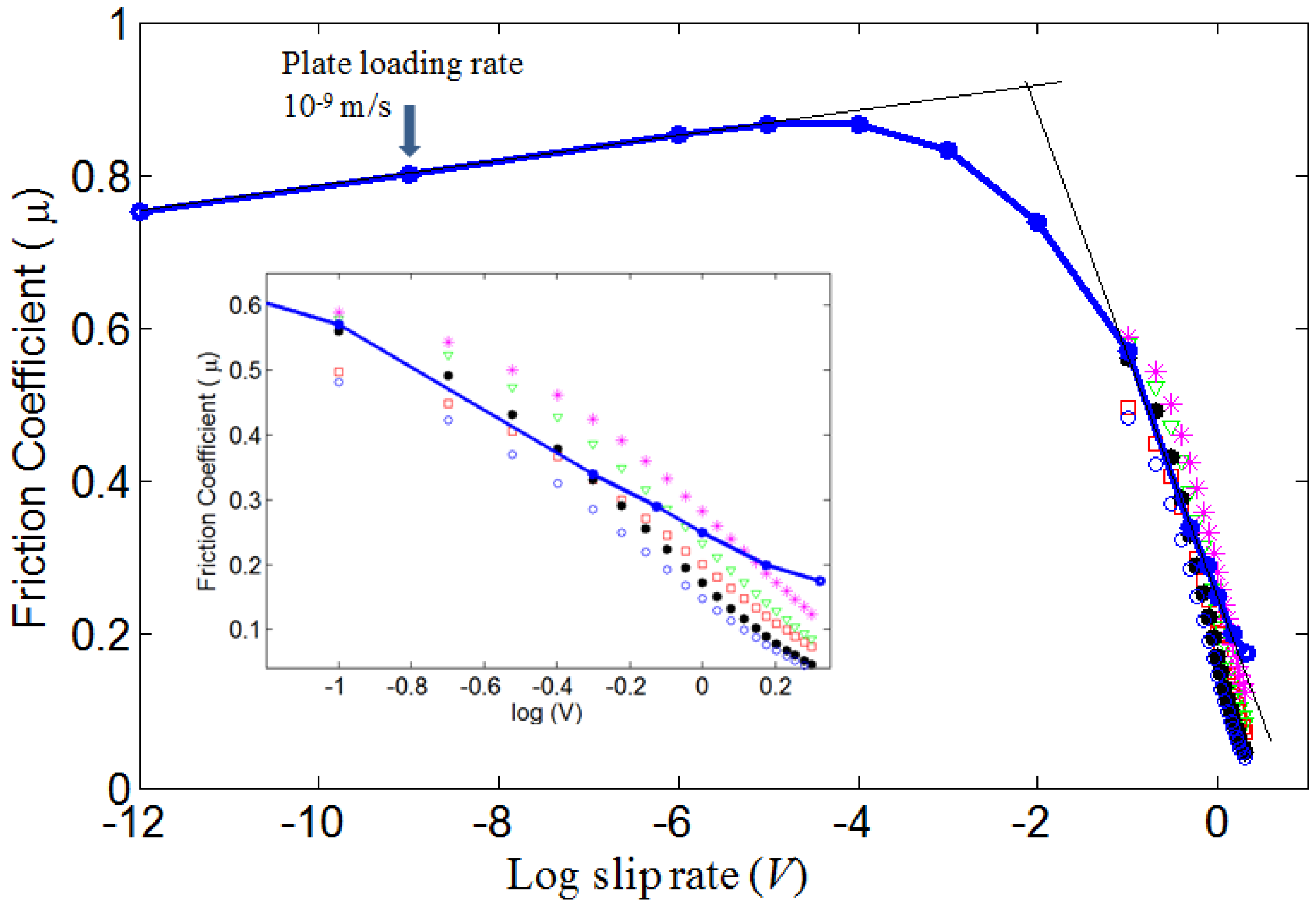

Fig. 4: Steady state friction coefficient as a function of slip rate (layer thickness = 1mm) at $T_b = 300\,\text{K}$. Blue curve represents the prediction of our model based on Eqn. (9) for the grain contact strength. Scattered points represent the sweep of experimental data from Sone and Shimamoto [2009]. Different colors correspond to different values for the fitting parameters in Sone and Shimamoto [2009].

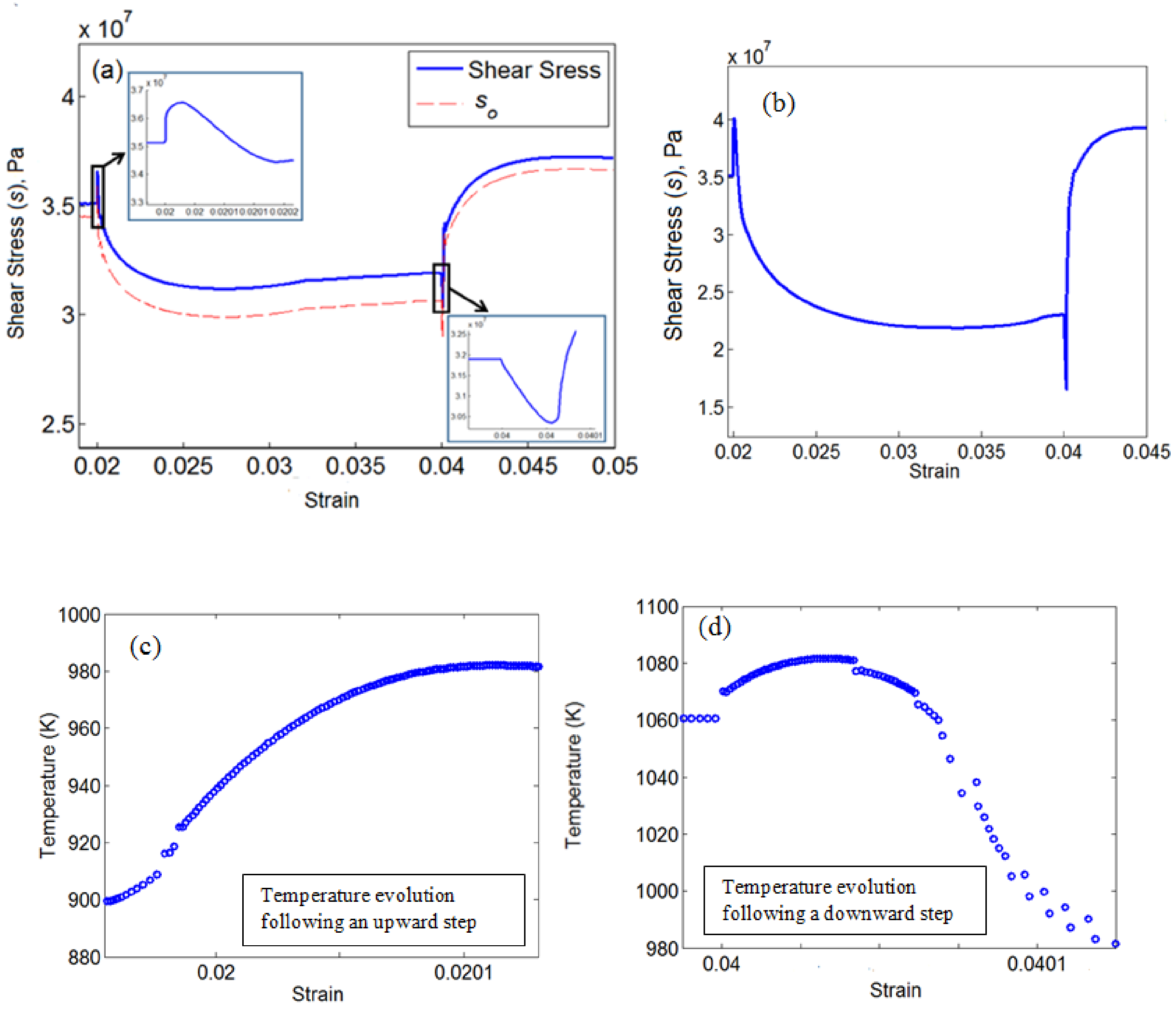

Fig. 5: Results for a velocity stepping numerical experiment. (a) Evolution of shear stress and minimum flow stress $s_o$. In the upward step, the strain rate is doubled. After steady state is reached the strain rate was reduced to its original value. Inserts show a magnified plot for the variation of shear stress immediately following the step. Steady state stress is reached after the downward step at strains greater than 0.05 (not shown here). (b) Evolution of shear stress in a pair of strain rate stepping experiment in which the strain rate ration is 10 (0.1) in the upward (downward) step (respectively) .(c) Evolution of contact temperature immediately following an upward step in velocity corresponding to Fig.(5a). (d) Evolution of contact temperature immediately following a downward step in velocity corresponding to Fig. (5a).

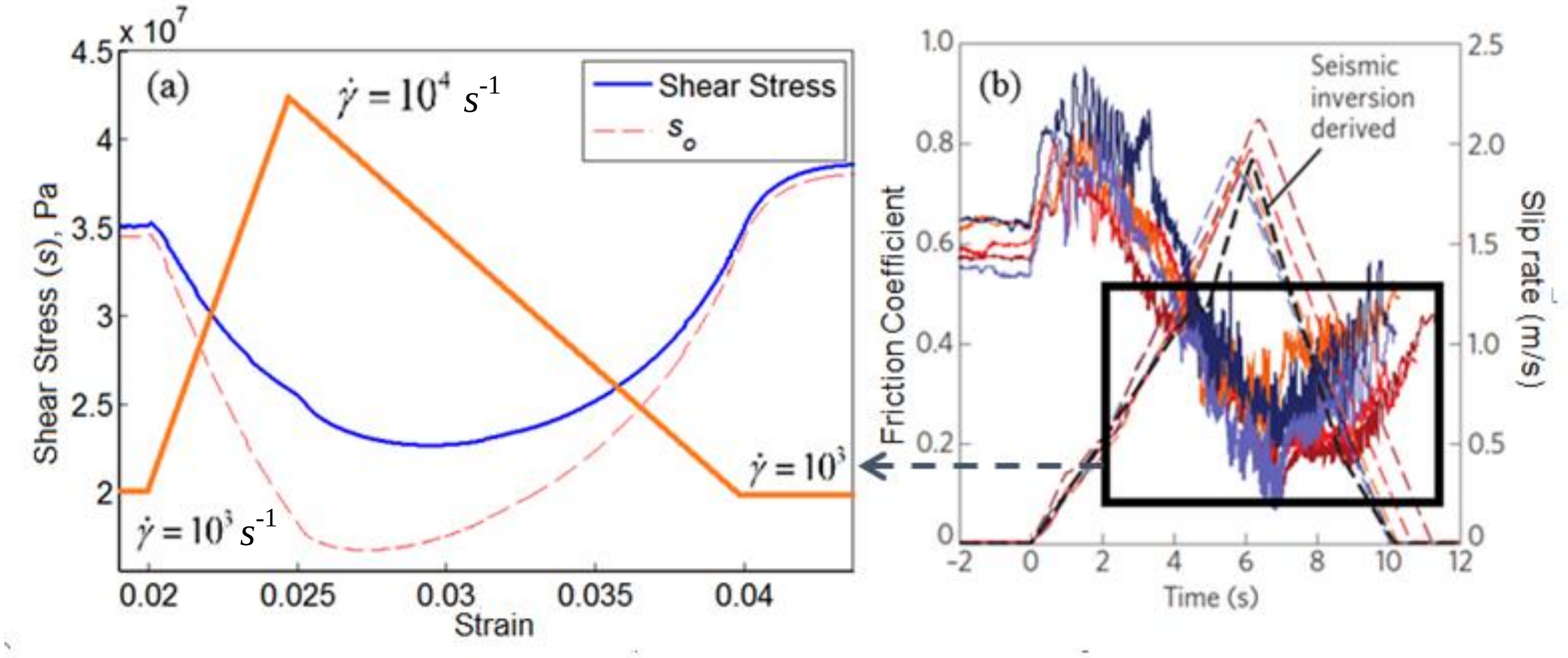

Fig. 6: Response to linear changes in imposed strain rates. (a) The model predicts gradual weakening followed by gradual strengthening as the velocity is ramped up then down. The brown solid line represents the velocity ramp (b) Experimental observations from Sone and Shimamoto [2009] (reprinted with permission) showing qualitatively similar behavior for the region within the black rectangle.

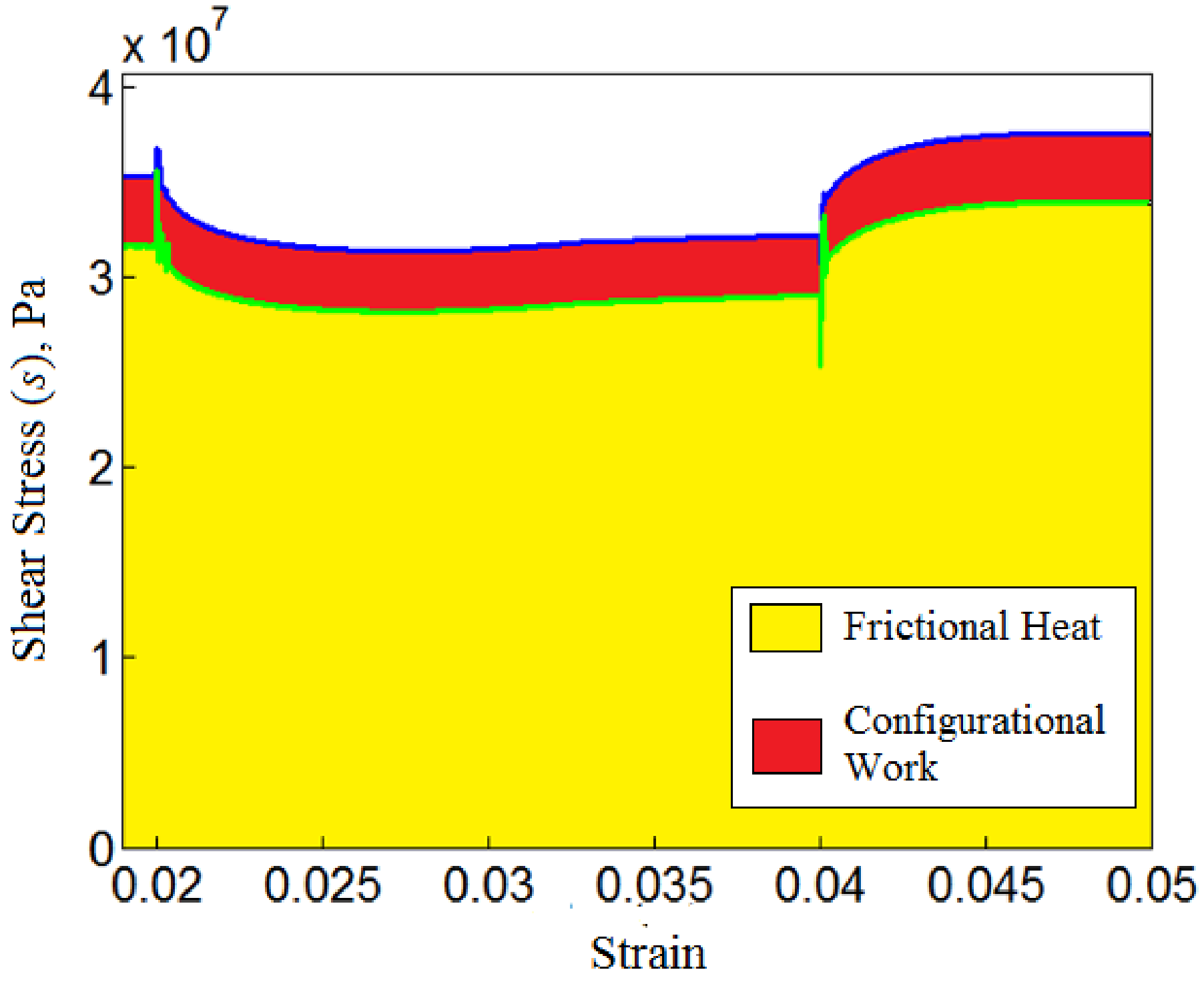

Fig. 7: Energy partitioning between frictional (yellow) and configurational (red) components. The blue line represents the macroscopic shear stress. The green line represents the fraction of the shear stress that is contributing to heat dissipation. For small strain, as shown here, the configurational energy is approximately 10% of the total energy budget. As slip further accumulates, the effective temperature evolves towards its steady state value and the fraction of energy consumed in increasing local disorder decreases. At steady state, all the external work is dissipated as heat.